\documentclass[aps,prd,preprint,floatfix,nofootinbib,showpacs]{revtex4-1}
\pdfoutput=1
\usepackage{graphicx,color,epstopdf,amsmath}
\usepackage{hyperref}

\newcommand{\BM}[1]{{\mbox{\boldmath{$#1$}}}}
\newcommand{\fr}[2]{{\hbox{$ #1 \over #2 $}}}

\begin{document}
{\small
\begin{flushright}
IPMU17-0062
\end{flushright} }
\title{Interpreting the 3 TeV $WH$ Resonance as a $W'$ boson}
\author{ Kingman Cheung$^{1,2,3}$, Wai-Yee Keung$^{4,5,1}$, Chih-Ting Lu$^{2}$,
Po-Yan Tseng$^{6}$}
\affiliation{
$^1$ Physics Division, National Center for Theoretical Sciences, Hsinchu,
Taiwan \\
$^2$Department of Physics, National Tsing Hua University, 
Hsinchu 300, Taiwan \\
$^3$Division of Quantum Phases \& Devices, School of Physics, 
Konkuk University, Seoul 143-701, Republic of Korea \\
$^4$Department of Physics, University of Illinois at Chicago, IL 60607, USA \\
$^5$Department of Physics and Astronomy, Northwestern University, 
Evanston, IL 60208, USA\\
$^6$Kavli IPMU (WPI), UTIAS, The University of Tokyo, Kashiwa, Chiba 277-8583, Japan
}

\renewcommand{\thefootnote}{\arabic{footnote}}
\date{\today}

\begin{abstract}
Motivated by a local $3.2-3.4$ sigma resonance in $WH$ and $ZH$ in the ATLAS
Run 2 data, we attempt to interpret the excess in terms of a $W'$ boson
in a $SU(2)_1 \times SU(2)_2 \times U(1)_X$ model. We stretch the 
deviation from the alignment limit of the Equivalence Theorem, so as
to maximize $WH$ production while keeping the $WZ$ production rate below the
experimental limit. We found a viable though small region of parameter
space that satisfies all existing constraints on $W' \to jj, t \bar b, WZ$,
as well as the precision Higgs data.
The cross section of $W' \to WH$ that we obtain is about 
$5-6$ fb.
\end{abstract}

\maketitle

\section{Introduction}
Recently, the ATLAS Collaboration \cite{atlas} reported an experimental
anomaly in $WH$ or $ZH$ production in the $q\bar q b\bar b$ final state
at $\sqrt{s} = 13$ TeV  with an apparent excess
at around 3 TeV resonance mass region.  
Note that CMS also searched for 
the same channels \cite{cms}. Though they did not
claim observing anything peculiar, we can see that there is a visible peak 
of more than $2\sigma$ at around 2.7 TeV.  
Currently, the CMS observation does not support the 3 TeV excess of ATLAS 
base on the narrow width resonance analysis.
The broad width analysis has not been fully studied, 
and so it is hard to make conclusion for broad width resonance case.
We shall focus on interpreting
the ATLAS result while we emphasize that the CMS result does not falsify
the ATLAS result.
The excessive cross section is roughly \cite{atlas} 
(which is estimated from the 95\% CL upper limits on the cross section curves)
\begin{equation} 
\label{first}
  \sigma (pp \to W' \to WH) \times B(H \to b\bar b) 
 \approx 5  \pm 1.5 \;{\rm fb} \;.
\end{equation}
A similar excess was seen in $ZH$ production.
The local excesses are at about $3.2 -3.4 \sigma$ for both $WH$ and $ZH$
channels at around 3 TeV, 
while the global significance is about $2.2\sigma$.
Nevertheless, the boosted hadronic decays of $W$ and $Z$ have substantial
overlap at about 60\% level, which means that it is difficult to
differentiate between the $W$ and $Z$ bosons. In the following, we
focus on the excess interpreted as a 3 TeV $WH$ resonance.

We attempt to interpret that there is a 3 TeV spin-1 resonance $W'$ 
that decays
into $WH$. The $W'$ can arise from a number of extended symmetric
models, e.g., $SU(2)_1 \times SU(2)_2 \times U(1)_X$ 
\cite{Dobrescu:2015yba,221}.
With an additional $SU(2)$ symmetry, which is broken at the multi-TeV scale, 
there will be extra $W'$ and $Z'$ bosons, whose masses may be similar
or differ depending on the symmetry-breaking pattern. Then the decay
$W' \to WH$ can explain the excess with a resonance structure. Similarly,
the $Z' \to ZH$ can explain the excess in $ZH$ production.  Here we 
focus on the $WH$ channel.

The $W'$ boson couples to the right-handed fermions with a strength 
$g_R$, independent of the left-handed weak coupling.
The $W'$ boson can then be produced via $q\bar q'$ annihilation. 
The $W'$ boson can mix with the standard model (SM) $W$ boson
via a mixing angle, say $\sin\phi_w$, so that the $W'$ boson can decay 
into $WZ$ and $WH$ with a mixing-angle suppression, and right-handed fermions.
Previously, there was the 2 TeV $WZ$ and $WW$ anomaly which motivated 
a lot of phenomenological activities. One of the constraints
was the $WH$ constraint because the Equivalence theorem (ET) states that 
$\Gamma(W' \to WZ) \approx \Gamma( W' \to WH)$ in the heavy $W'$ limit
\cite{previous}.
In the model that we are considering, it is indeed true in the alignment limit
$\beta \to \pi/2+\alpha$.
Here we attempt to explore how much we can 
deviate from the alignment limit so that the $WH$ channel can be enhanced
while suppressing the $WZ$, thus satisfying the constraint from 
$WZ$ \cite{ATLAS:2016yqq, ATLAS:2016cwq, ATLAS:2016npe, CMS:2016mwi, CMS:2016pfl},
dijet \cite{ATLAS:dijet, CMS:dijet}, 
and precision Higgs data \cite{atlas-2hdm}.
\footnote
{The leptonic constraint on $W'$ and $Z'$ are so strong that 
we opt for the leptophobic nature for the $W'$ and $Z'$ bosons.
}

The organization of this note is as follows. 
In the next section, we describe 
the $SU(2)_1 \times SU(2)_2 \times U(1)_X$ model that we consider in this
work.  In Sec. III, we demonstrate the deviation from the alignment limit.
In Sec. IV, we discuss all the relevant constraints.  We present 
the results in Sec. V, and conclude and comment in Sec. VI.

\section{The $SU(2)_1 \times SU(2)_2 \times U(1)_X$ model}

We follow \cite{Dobrescu:2015yba,221} a 
renormalizable model based on the $SU(2)_L \times SU(2)_R \times U(1)_{B-L}$
symmetry. In addition to the SM fermions and gauge bosons, this model
also contains new gauge bosons $ W' $, $ Z' $, the right-handed
neutrinos $ N_R $, and also some
extra scalars from the extended Higgs sector: a complex $ SU(2)_R $
triplet $ T $ and a complex $ SU(2)_L \times SU(2)_R $ bidoublet $
\Sigma $. We summarize the particle contents and gauge charges in
Table~\ref{tab:model} of this $SU(2)_L \times SU(2)_R \times
U(1)_{B-L}$ Model.

\begin{table}[h!]
\caption{\small  \label{tab:model}
The particle contents and gauge charges of the $SU(2)_L \times SU(2)_R \times U(1)_{B-L}$ Model \cite{Dobrescu:2015yba}.
}
\vspace{1.0mm}
\begin{ruledtabular}
 \begin{tabular}{ l c c c }
Fields & $ SU(2)_{L} $ & $ SU(2)_R $ & $ U(1)_{B-L} $ \\ \hline 
$ (u_L ,d_L) $ & 2 & 1 & +1/3 \\
$ (u_R ,d_R) $ & 1 & 2 & +1/3 \\
$ (\nu_L ,l_L) $ & 2 & 1 & -1 \\
$ (N_R ,l_R) $ & 1 & 2 & -1 \\
$ \Sigma $ & 2 & 2 & 0 \\
$ T $ & 1 & 3 & +2 \\
\end{tabular}
\end{ruledtabular}
\end{table}

We focus on the extended Higgs sector to study the mass and mixing 
of new gauge bosons $ W' $, $ Z' $. There are two steps of symmetry 
breaking from two 
sets of complex scalar fields, separately.
First, the $ SU(2)_R $ triplet scalar 
$ T =(T^{++}, T^+, T^0)  $ 
breaks $ SU(2)_R \times U(1)_{B-L} $ to $ U(1)_Y $ by  acquiring  a large 
vacuum-expectation value (VEV)
at the multi-TeV scale.
\[
  \langle T \rangle = (0,0, u_T)^T   \ .
\]
The heavy masses of $W'$ and $Z'$ are set by $ u_T $.
Second, the $ SU(2)_L \times SU(2)_R $ bidoublet scalar, 
\begin{equation} 
  \Sigma  =
  \left( \begin{array}{cc} 
         \Phi^{0*}_1 & \Phi^{+}_2 \\
         -\Phi^{-}_1  & \Phi^0_2       \end{array}\right ) \; ,
\end{equation}
develops a VEV at the electroweak scale
$v=(v_1^2+v_2^2)^{1\over2} \approx 246$~GeV. 
\begin{equation}
  \langle \Sigma \rangle = 
\frac{1}{\sqrt{2}}
  \left( \begin{array}{cc} 
         v_1 & 0 \\
         0  & e^{i\alpha_\Sigma} v_2 \end{array} \right ) \;
=
\frac{v}{\sqrt{2}}
  \left( \begin{array}{cc} 
         \cos\beta & 0 \\
         0  & e^{i\alpha_\Sigma} \sin\beta \end{array} \right ) \; ,
\end{equation}
which further breaks $ SU(2)_L \times U(1)_Y $ to $ U(1)_Q $, where
$Q=T_3^L+T_3^R+\fr12(B-L)$.
The phase $ \alpha_\Sigma$ 
is  CP-violating, and  we do not include its effects in
this work. 
The ratio $ \tan\beta=v_2/v_1 $  of two VEV's
follows the same notation as two-Higgs-doublet models (2HDM). This
symmetry breaking induces a small mixing between the charged gauge bosons.

Explicitly, the field content of $\Sigma$ is given by
\begin{eqnarray}
\Phi^0_1 &=& \fr1{\sqrt{2}}[ v_1+(-H\sin\alpha +H'\cos\alpha -iA^0\sin\beta
                                                         +iG^0\cos\beta ) ]
\ , \nonumber \\
\Phi^0_2 &=& \fr1{\sqrt{2}}[ v_2+( H\cos\alpha +H'\sin\alpha +iA^0\cos\beta
                                                       +iG^0\sin\beta ) ]
\ , \nonumber \\
\Phi^+_1 &=& \cos\beta G^+ - \sin\beta H^+ \ , \nonumber \\
\Phi^+_2 &=&  \sin\beta G^+ + \cos\beta H^+ \ , \label{eq:phis}
\end{eqnarray}
with the $H$ being the observed 125 GeV Higgs boson, $H'$ the heavy Higgs boson,
$H^\pm$ the charged Higgs boson, $A$ the pseudoscalar Higgs boson, 
and $G^{\pm}$, $G^0$ the Nambu-Goldstone bosons.

We are interested in the energy scale $u_T$ much larger than the
electroweak scale $v$. Therefore, the scalar fields from the triplet $T$
are decoupled from the electroweak scale. At the energy scale lower than
$u_T$, the scalar sector only consists of the bidoublet $\Sigma$,
which is the same as the 2HDM with the doublet fields
$H^T_1=(\Phi^+_1,\Phi^0_1)^T$ and
$H^T_2=(\Phi^+_2,\Phi^0_2)^T$~\cite{Dobrescu:2015yba}.

The electrically-charged states, $ W^{\pm}_L $ and $ W^{\pm}_R $, of the
$ SU(2)_L $ and $ SU(2)_R $ symmetries will mix to form physical gauge bosons, 
$ W^{\pm} $ and $ W'^{\pm} $, 
\begin{equation}
 \left( \begin{array}{c} 
              W^{\pm} \\
              W'^{\pm} \end{array}  
  \right ) = 
 \left( \begin{array}{cc} 
        \cos\phi_w & \sin\phi_w \\
       -\sin\phi_w & \cos\phi_w  \end{array} \right ) \;
 \left( \begin{array}{c} 
               W^{\pm}_L \\
               W^{\pm}_R \end{array}  \right )  \;.
\end{equation}
The $ W^{\pm}_L - W^{\pm}_R $ mixing angle $ \phi_w $ satisfies
\begin{equation}
  \sin\phi_w = \frac{g_R}{g_L}\left(\frac{m_W}{m_{W'}}\right)^2 \sin2\beta\;,
\end{equation}
and the $ W $ and $ W' $ masses are given by
\begin{equation}
  m_W = \fr12 g_L v \ ,\;\;\;
  m_{W'} = g_R u_T\;,
\end{equation}
where $ g_L $ and $ g_R $ are the $SU(2)_L \times SU(2)_R$ gauge couplings.
We assume that the mass of the right-handed neutrino is heavier than the $W'$, 
such that the decay $W' \to l_R N_R$ is kinematically forbidden.

There are other possible decay modes for the $W'$ into other Higgs bosons
\cite{Dobrescu:2015yba} if they are kinematically allowed:
e.g.,
\[
  W'^+ \to H^+A,\;Z H^+,\; W^+ H',\; W^+ A,\; H^+ H,\; H^+ H' \;.
\]
Such decay widths depend on the mass parameters and are highly model dependent,
and so we treat the sum of these decay widths as a restricted variable
parameter denoted by $\Gamma^{\rm other}_{W'}$.

\section{Deviations from the Alignment limit}

In this section, we would derive the $W'WZ$ and $W'WH$ couplings in
this $SU(2)_L \times SU(2)_R \times U(1)_{B-L}$ model, using the
2HDM convention, by rewriting the bidoublet $\Sigma$ in
terms of two doublets $(\Phi^+_1,\Phi^0_1)^T$ and
$(\Phi^+_2,\Phi^0_2)^T$~\cite{Dobrescu:2015yba}.  The deviation from
the ET, $\Gamma(W' \to WZ)\neq \Gamma(W' \to WH)$,
can be realized, if the mixing angles $\alpha$ and $\beta$ in 2HDM
stays away from the alignment limit.  Or vice versus, the
ET is restored when $\beta \to \alpha+\pi/2$.

The mass mixing term between $W$ and $W'$ comes from the bidoublet and
is given by,
with the VEV's of the decomposed doublets denoted by
$v_1=v\cos\beta$ and $v_2=v\sin\beta$,
\begin{equation}
 m_{WW'}^2 = 2 \times \frac{g_R}{\sqrt{2}}  \frac{v_1}{\sqrt{2}}
\frac{g_L}{\sqrt{2}} \frac{v_2}{\sqrt{2}} =\frac{g_R g_L}{2} v_1 v_2 \  . 
\end{equation}
Note that the factor of 2 in front comes from two ways of matchings.
So the induced mixing is described by
\begin{equation}
 \left(\begin{array}{c} W  \\ W' \end{array}\right)
=\left(\begin{array}{cc} \cos\phi_w & -\sin\phi_w   \\ 
                        \sin\phi_w & \cos\phi_w       \end{array}\right)
\left(\begin{array}{c} W_L \\ W_R \end{array}\right)
\ ,\quad
\sin\phi_w \approx m_{WW'}^2/m_{W'}^2 
\ . \end{equation}
Similarly, there is mixing between $Z$ and $Z'$.
The mixing angle $\phi_w$ induces the coupling $W^\dagger W'Z$ 
from the gauge vertices 
$W_L^\dagger  W_L Z$ and $W_R^\dagger W_R Z$ of different strengths,
according to the SM pattern $T^L_3-Q\sin^2\theta_W$.
The two contributions sum up to
$$ \frac{g_L}{\cos\theta_W}\sin\phi_w
               \left[ -(0-\sin^2\theta_W) +(1-\sin^2\theta_W)\right]
  = \frac{g_L}{\cos\theta_W}\sin\phi_w   \ .$$
\begin{equation} (\hbox{coupling of } W'^\dagger W Z)\equiv g_{W'WZ}
  =\frac12\frac{g_L^2 g_R v^2}{ m_{W'}^2\cos\theta_W} 
      \sin\beta\cos\beta 
  = g_R \frac{m_W m_Z}{m_{W'}^2} \sin2\beta 
\ .  \label{eq:WPWZcoupling} \end{equation}
However, the leading vertex $W'^\dagger W H$ is given {\it not} explicitly
from the mixing, but derived  by the following steps,
\begin{equation}
 (\hbox{coupling of } W'^\dagger W H)\equiv g_{W'WH}
=\frac{\partial m_{WW'}^2 }{\partial (v_1/\sqrt2)} 
                              \frac{\partial \Phi_1^0 }{\partial H}+
\frac{\partial m_{WW'}^2 }{\partial (v_2/\sqrt2)} 
                              \frac{\partial \Phi_2^0 }{\partial H} 
\ . \end{equation}
Therefore,
\begin{equation}
 g_{W'WH} =\frac{g_L g_R v}{2}
      \left[-\frac{v_2}{v}\sin\alpha 
            +\frac{v_1}{v}\cos\alpha \right] 
    = g_R m_W \cos(\alpha+\beta) 
\ . \end{equation}
Similarly, the Goldstone boson $G^0$, 
associated with $Z$, also accompanies with
$H$. 
\begin{equation}
 (\hbox{coupling of } W'^\dagger W G^0)  \equiv g_{W'W G^0} =
\frac{\partial m_{WW'}^2 }{\partial (v_1/\sqrt2)}
                              \frac{\partial \Phi_1^0 }{\partial G^0}+
\frac{\partial m_{WW'}^2 }{\partial v_2/\sqrt2} 
                          \frac{\partial \Phi_2^0 }{\partial G^0}
\ \ . \end{equation}
\begin{equation}
 g_{W'W G^0} =\frac{g_L g_R v}{2} i
      \left[\frac{v_2}{v}\cos\beta
            +\frac{v_1}{v}\sin\beta \right]
    =i\frac{g_L g_R v}{2}  \left[ 2\sin\beta\cos\beta \right]
    =i g_R m_W \sin2\beta
\ . \end{equation}
In summary,
{$ g_{W'W G^0} =i g_R m_W \sin2\beta$},
and
{$ g_{W'W H} = g_R m_W \cos(\alpha+\beta)$}. 
Thus, we obtained the decay widths for $W' \to WZ$ and $W' \to WH$ in the limit $m_{W'} \gg m_{W,Z,H}$.
\begin{eqnarray}
\Gamma(W' \to WZ) &\simeq & \frac{g^2_R}{192 \pi} m_{W'}\, \sin^2 2\beta \,, \nonumber \\
\Gamma(W' \to WH) &\simeq & \frac{g^2_R}{192 \pi} m_{W'}\, \cos^2 (\alpha+\beta) \,. \label{eq:widths}
\end{eqnarray}

In the alignment limit, $\alpha \to \beta-\fr\pi2$, the two widths above 
become equal. As ET identifies $G^0$ with the
longitudinal $Z$,  we expect the relations, 
\begin{equation}
\Gamma(W'\to WZ) \approx \Gamma(W'\to WG^0) 
         \approx \Gamma(W'\to WH) 
\hbox{ as } \alpha \to \beta-\fr\pi2
\ .\end{equation}

We are going to illustrate the operation of  the ET.
The longitudinal $W^+$ is  identified with  
$G^+=\cos\beta \Phi^+_1 + \sin\beta \Phi^+_2    $ in 
Eq.(\ref{eq:phis}).
The action of $W'$  moves entries within the same row 
in the $2 \times 2$ matrix form of the bidoublet.
Therefore the amplitude 
$$ {\cal M}(W'\to G^+(p^+) G^0(p^0) ) =  \frac{g_R}{\sqrt2} 
\frac1{\sqrt2} (\cos\beta\sin\beta + \cos\beta\sin\beta)
\  (p^+ -p^0)\cdot \epsilon'   \ .$$
\begin{equation} {\cal M}(W'\to G^+(p^+) G^0(p^0) )  
=\frac{g_R}{2}\sin2\beta \  (p^+ -p^0)\cdot \epsilon'
\ .  \end{equation}
The factor $(p^+ -p^0)$ corresponds to the Feynman amplitude for the
convective current, which is contracted with the polarization vector
$\epsilon'$ of $W'$. The above amplitude should give the same width 
$\Gamma(W'\to W G^0)$. Indeed it is because 
$\fr12(p^+ -p^0)\cdot \epsilon' = p^+\cdot \epsilon' 
\approx m_W \epsilon^+_L\cdot \epsilon'$.

On the other hand, we can start from the tri-gauge coupling of the 
anti-symmetric Lorentz form,
$$ (p^+-p^0)\cdot \epsilon' (\epsilon^+\cdot\epsilon^0)
  +(p^0+P)\cdot \epsilon^+ (\epsilon^0\cdot\epsilon')
  +(-P-p^+)\cdot \epsilon^0 (\epsilon'\cdot\epsilon^+)   $$
$$ \qquad 
= (2 p^+ \cdot \epsilon') (\epsilon^+ \cdot\epsilon^0)
 + (2 p^0 \cdot \epsilon^+) (\epsilon^0\cdot\epsilon')
  -(2 p^+\cdot \epsilon^0) (\epsilon'\cdot\epsilon^+)  \ .$$
Now using the ET, we concentrate at the longitudinally polarized 
$W$ of $\epsilon^+\approx p^+/m_W$ and $Z$ of $\epsilon^0\approx p^0/m_Z$. 
Up to an over factor $\fr1{m_W m_Z}$, we obtain
$$  (2 p^+ \cdot \epsilon') (p^+ \cdot p^0)
  + (2 p^0 \cdot p ^+) (p^0\cdot\epsilon')
  - (2 p^+\cdot p^0) (\epsilon'\cdot p^+)    $$
$$ \qquad =  (2 p^0 \cdot p ^+) (p^0\cdot\epsilon')
= m_{W'}^2 (p^0\cdot\epsilon')
\ .$$
Therefore, the longitudinal amplitude from Eq.(\ref{eq:WPWZcoupling}) 
agrees with the other calculation based on $G^+G^0$.
$$ {\cal M}(W'\to WZ)
= g_R\sin2\beta p^0\cdot \epsilon'
= \frac{1}{2} g_R \sin2\beta m_{W'} \hat{\BM{p}}\BM{\cdot\epsilon'} 
= \frac{1}{2} g_R m_{W'} \cos\theta \sin2\beta  \ .$$
Integrating out the angular parameter $\theta$, the decay width is
\begin{equation}
 \Gamma(W'\to WZ)=\frac{1}{2m_{W'}} \int |{\cal M}(W'\to WZ)|^2 
\frac{d\cos\theta}{2} \frac{1}{8\pi}   
=\frac{g_R^2\sin^22\beta}{192\pi}m_{W'}  
\ , \end{equation}
which is in agreement with Eq.(\ref{eq:widths}).

Following the similar method, 
we can verify the coupling of $WWH$ in this model by using
$$
m^2_{WW}=\frac{g^2_Lv^2}{4}=\frac{g^2_L}{4}(v^2_1+v^2_2)\,.
$$
Then the coupling of $WWH$ is 
\begin{equation}
g_{WWH}
=\frac{\partial m_{WW}^2 }{\partial (v_1/\sqrt2)} 
                              \frac{\partial \Phi_1^0 }{\partial H}+
\frac{\partial m_{WW}^2 }{\partial (v_2/\sqrt2)} 
                              \frac{\partial \Phi_2^0 }{\partial H}=g_L m_W \sin(\beta-\alpha)\,.
\end{equation}
In the alignment limit, $\beta \to \alpha + \pi/2$, 
the $WWH$ coupling goes back to the SM Higgs-gauge boson coupling.

Gauge-boson and fermonic couplings of the 125 GeV Higgs boson are now
well measured by ATLAS and CMS, especially, the couplings to the 
massive gauge bosons. 
The deviations from the SM values shall be less than about 10\%, 
i.e $|\sin(\beta-\alpha)|\gtrsim0.9$. 
That implies the allowed range of $|\cos(\beta-\alpha)|\lesssim 0.44$. 
Weaker limits for the couplings to up- and down-type quarks from Higgs 
precession data 
also dictate the $\alpha$ and $\beta$'s parameter region.
Therefore, in this model framework, 
the Higgs precision data would set the boundary on the deviation 
from the alignment limit, 
and thus restrict the ratio of $\Gamma(W' \to WZ)$ and $\Gamma(W' \to WH)$.

The robust and detailed allowed region of $\alpha$ and $\beta$ from 
Higgs precision data 
depends on different types of 2HDM's. 
For the allowed parameter region, we refer to Ref.~\cite{atlas-2hdm}, 
where Type-I, -II, Lepton-specific, and Flipped 2HDMs have been studied.
The universal feature from their results, 
in the small $\tan \beta\simeq 0.1$ region, 
the allowed $\cos(\beta-\alpha)$ is close to the alignment limit, 
i.e $|\cos(\beta-\alpha)|\lesssim 0.05$. 
This is because the universal up-type quark Yukawa coupling among the 2HDMs 
is enhanced by factor $1/\sin\beta$.
For $\tan\beta\gtrsim 2$ region, only the Type-I case allows more 
dramatic deviation from the alignment limit. 
For instance, taking $\tan\beta = 2.5$, 
the allowed range from Higgs precision data is 
$ -0.37 < \cos(\beta-\alpha) < 0.42$.
Because only in Type-I case, 
all the up-, down-quark and leptonic Yukawa couplings deviate from SM values 
by the same factor $(\cos\alpha/\sin\beta)$, 
such that larger $\tan\beta$ would not enhance any of these couplings, 
and they are therefore less constrained by Higgs precision data. 
We shall use the results of Type-I 2HDM obtained in Ref.~\cite{atlas-2hdm}
to restrict the parameter of our model.

\section{Constraints from existing data}

Recently, both ATLAS and CMS collaborations have published their $W'$
searches with different decay channels, including fermionic final
states $ l^{\pm}\nu $ ~\cite{ATLAS:lnu}, dijet~\cite{ATLAS:dijet,
  CMS:dijet}, $ t b $ ~\cite{CMS:2016wqa}, and also bosonic final
states $ W^{\pm}Z $ ~\cite{ATLAS:2016yqq, ATLAS:2016cwq,
  ATLAS:2016npe, CMS:2016mwi, CMS:2016pfl} at 13 TeV.  Here we list
all the constraints from these searches in
Table~\ref{tab:constraint}. Here $j$ includes all light flavors, $ l $
includes ($e, \mu$) and $\nu$ includes ($\nu_{e}, \nu_{\mu}, \nu_\tau$).
 Finally, $ J $ means large-$R$ jets ($W$ jet or $Z$ jet).

\begin{table}[h!]
\caption{\small  \label{tab:constraint}
Different decay mode searches at 13 TeV of $W'$ with $ m_{W'}\sim 3 $ TeV for both ATLAS and CMS constraints.
Here $j$ includes all light flavors, $ l $ and $ \nu $  include ($e, \mu$) and ($\nu_{e}, \nu_{\mu}$), separately. Finally, $ J $ means large-$R$ jets ($W$ jet or $Z$ jet).
}
\vspace{1.0mm}
\begin{ruledtabular}
 \begin{tabular}{ l c c c }
 & Process & Upper Bound & Ref.  \\ \hline
ATLAS & $ p p\rightarrow W^{\prime\pm}\rightarrow l^{\pm}\nu $ & $ \leq 0.243 $ (fb) & \cite{ATLAS:lnu} \\
ATLAS & $ p p\rightarrow W^{\prime\pm}\rightarrow j j' $ & $ \leq 69.5 $ (fb) & \cite{ATLAS:dijet} \\
CMS & $ p p\rightarrow W^{\prime\pm}\rightarrow j j' $ & $ \leq 41.7 $ (fb) & \cite{CMS:dijet} \\
CMS ($ t b\rightarrow l^{\pm}\nu b b $) & $ p p\rightarrow W^{\prime\pm}\rightarrow t b $ & $ \leq 84.4 $ (fb) & \cite{CMS:2016wqa} \\
ATLAS ($ W^{\pm}Z\rightarrow J J $) & $ p p\rightarrow W^{\prime\pm}\rightarrow W^{\pm}Z $ & $ \leq 3.0 $ (fb) & \cite{ATLAS:2016yqq} \\
ATLAS ($ W^{\pm}Z\rightarrow l^{\pm}\nu q q' $) & $ p p\rightarrow W^{\prime\pm}\rightarrow W^{\pm}Z $ & $ \leq 5.5 $ (fb) & \cite{ATLAS:2016cwq} \\
ATLAS ($ W^{\pm}Z\rightarrow q q' l^+ l^- $) & $ p p\rightarrow W^{\prime\pm}\rightarrow W^{\pm}Z $ & $ \leq 10.4 $ (fb) & \cite{ATLAS:2016npe} \\
ATLAS ($ W^{\pm}Z\rightarrow q q'\nu\nu $) & $ p p\rightarrow W^{\prime\pm}\rightarrow W^{\pm}Z $ & $ \leq 3.0 $ (fb) & \cite{ATLAS:2016npe} \\
CMS ($ W^{\pm}Z\rightarrow J J $) & $ p p\rightarrow W^{\prime\pm}\rightarrow W^{\pm}Z $ & $ \leq 3.2 $ (fb) & \cite{CMS:2016mwi} \\
CMS ($ W^{\pm}Z\rightarrow l^{\pm}\nu q q $) & $ p p\rightarrow W^{\prime\pm}\rightarrow W^{\pm}Z $ & $ \leq 7.0 $ (fb) & \cite{CMS:2016pfl} \\
\end{tabular}
\end{ruledtabular}
\end{table}  
  
As we can see from Table~\ref{tab:constraint} that the strongest
constraint comes from $ W^{\prime\pm}\rightarrow l^{\pm}\nu $ searches,
but here we choose the leptophobic version of the model such that this
constraint will not cause serious effects on our results. 
On the other hand, 
the dijet constraints from both the ATLAS and CMS analyses rely on
the acceptance ($A$) and the width-to-mass ratio ($\Gamma /M$)
effects. 
Note that the dijet limits
  quoted in Table~\ref{tab:constraint} are only for the narrow-width
  resonance scenario. 
Here we follow their analyses by using $A=0.4 (0.6)$ for
ATLAS \cite{ATLAS:dijet} (CMS \cite{CMS:dijet}) analyses and the 
width-to-mass ratio effects are from Table 2 in \cite{ATLAS:dijet} for ATLAS
analysis and Table 4 in for CMS analysis \cite{Khachatryan:2015sja} to
rescale in our case.
\footnote{ Since we do not find the width-to-mass ratio effects for 
  $W^{\prime\pm}\rightarrow t b $ and $ W^{\prime\pm}\rightarrow
  W^{\pm}Z $ for either ATLAS or CMS analysis, we therefore conservatively
  use the original constraints of their publications with the narrow width
  approximation analysis.  }  

Another set of constraints come from the precision Higgs boson data,
including the gauge-Higgs couplings, Yukawa couplings, and the
$H\gamma\gamma$ and $Hgg$ factors.  In 2HDMs, such constraints can be
recast in terms of $\tan\beta$ and $\cos(\beta - \alpha)$. The excluded
region in the parameter space of Type-I 2HDM is shown explicitly 
in the upper-left panel in Fig.~\ref{f1} \cite{atlas-2hdm}. 

\section{Results}

\begin{figure}[t!]
\centering
\includegraphics[height=3.2in,angle=-90]{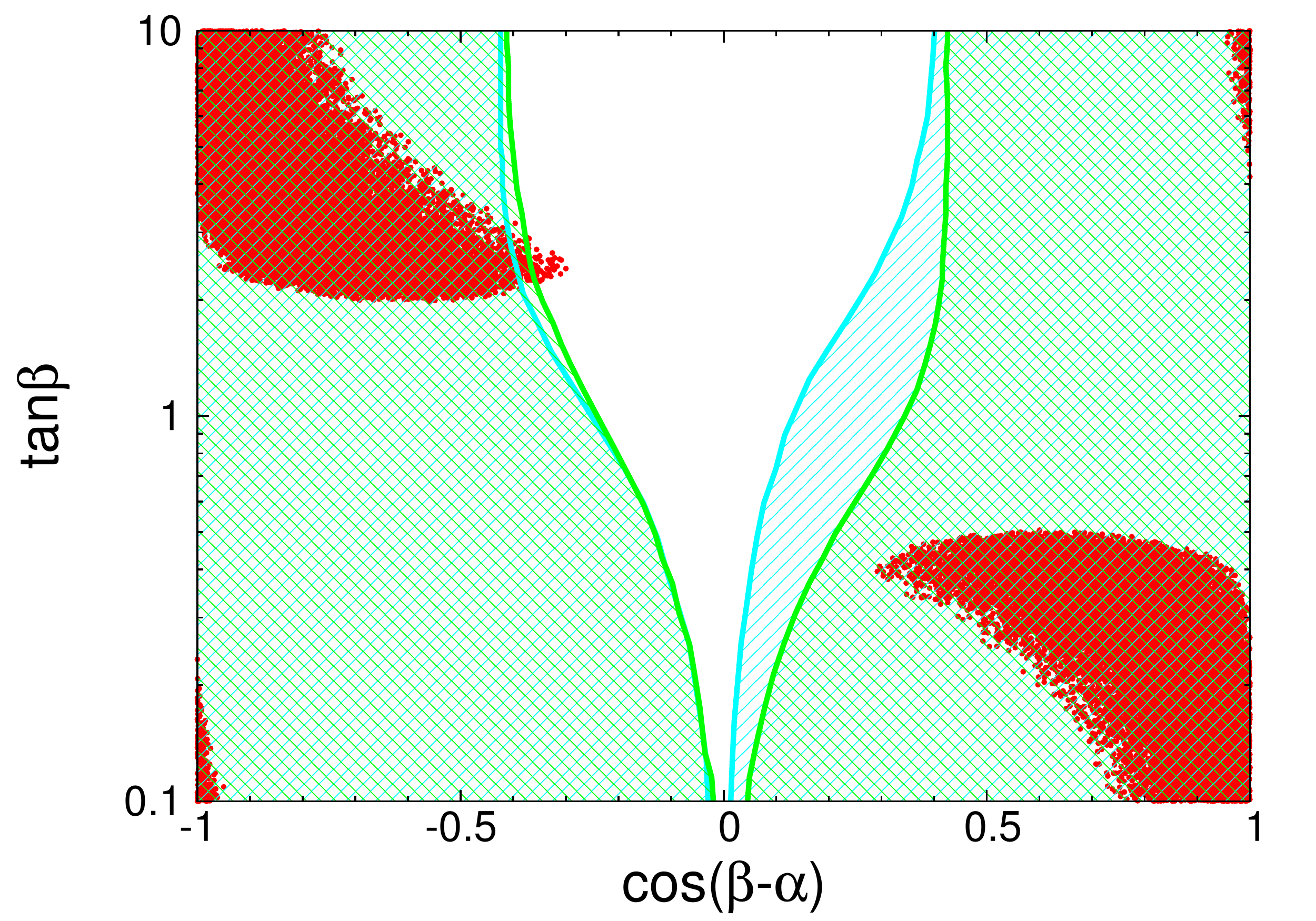}
\includegraphics[height=3.2in,angle=-90]{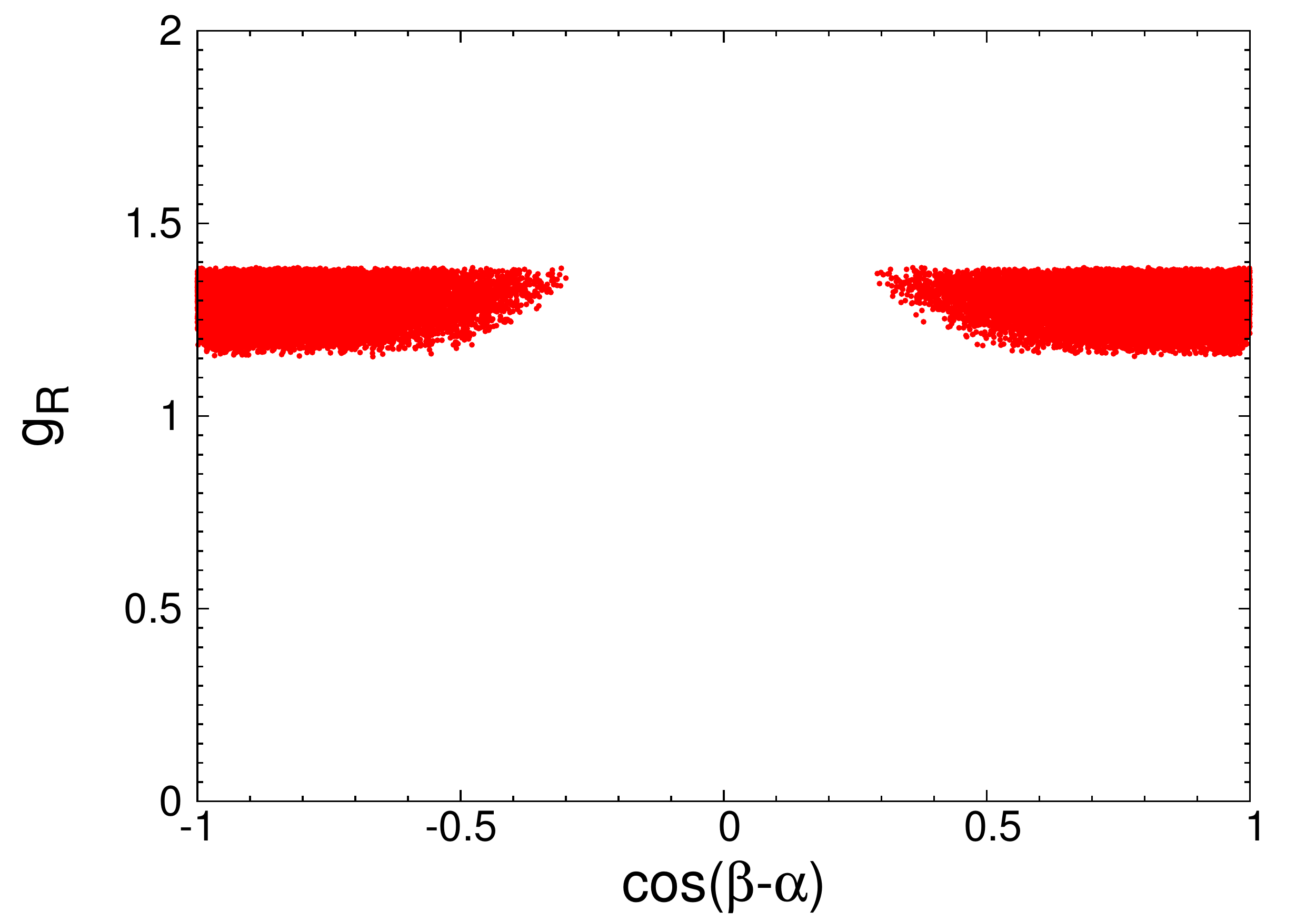}
\includegraphics[height=3.2in,angle=-90]{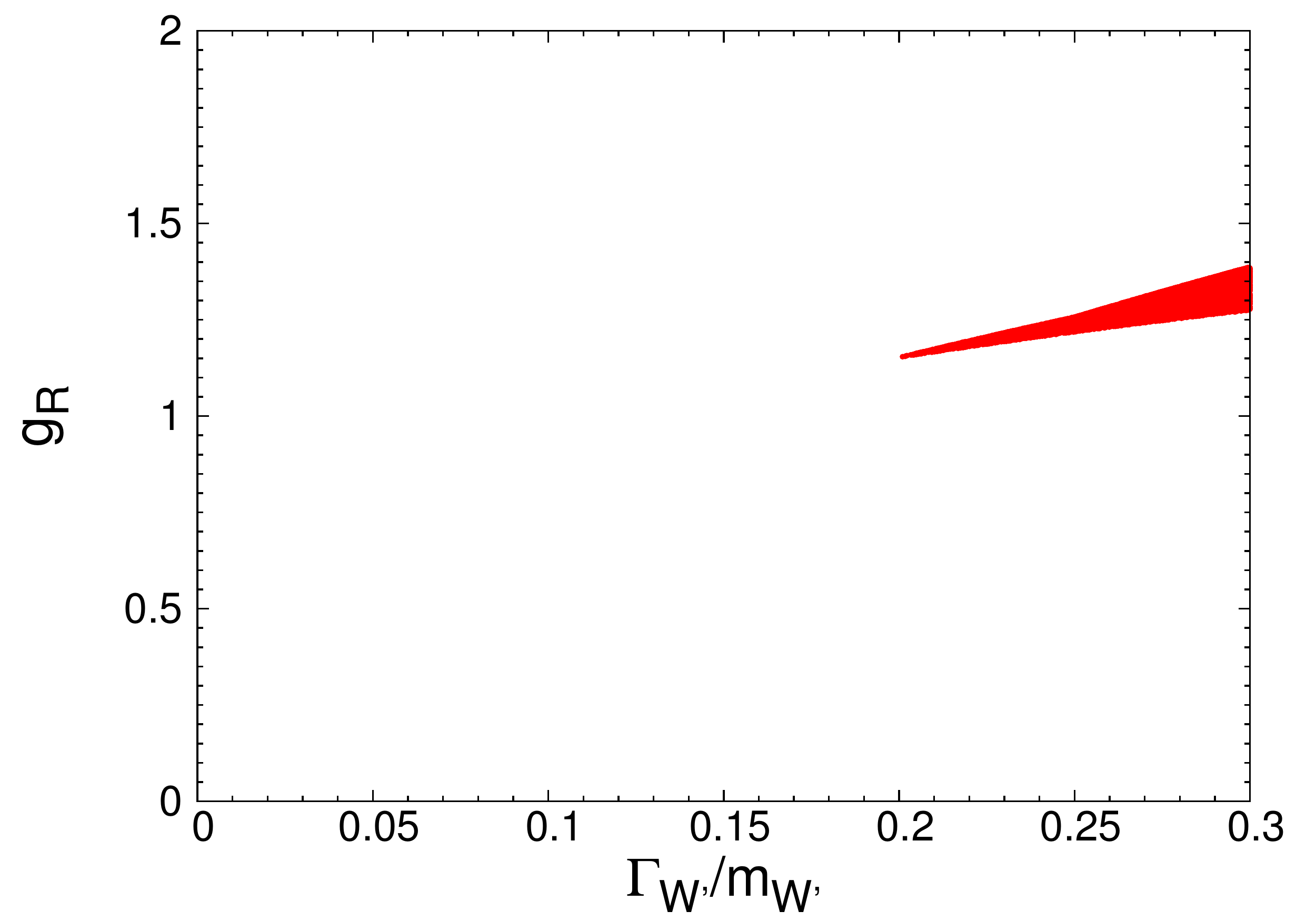}
\includegraphics[height=3.2in,angle=-90]{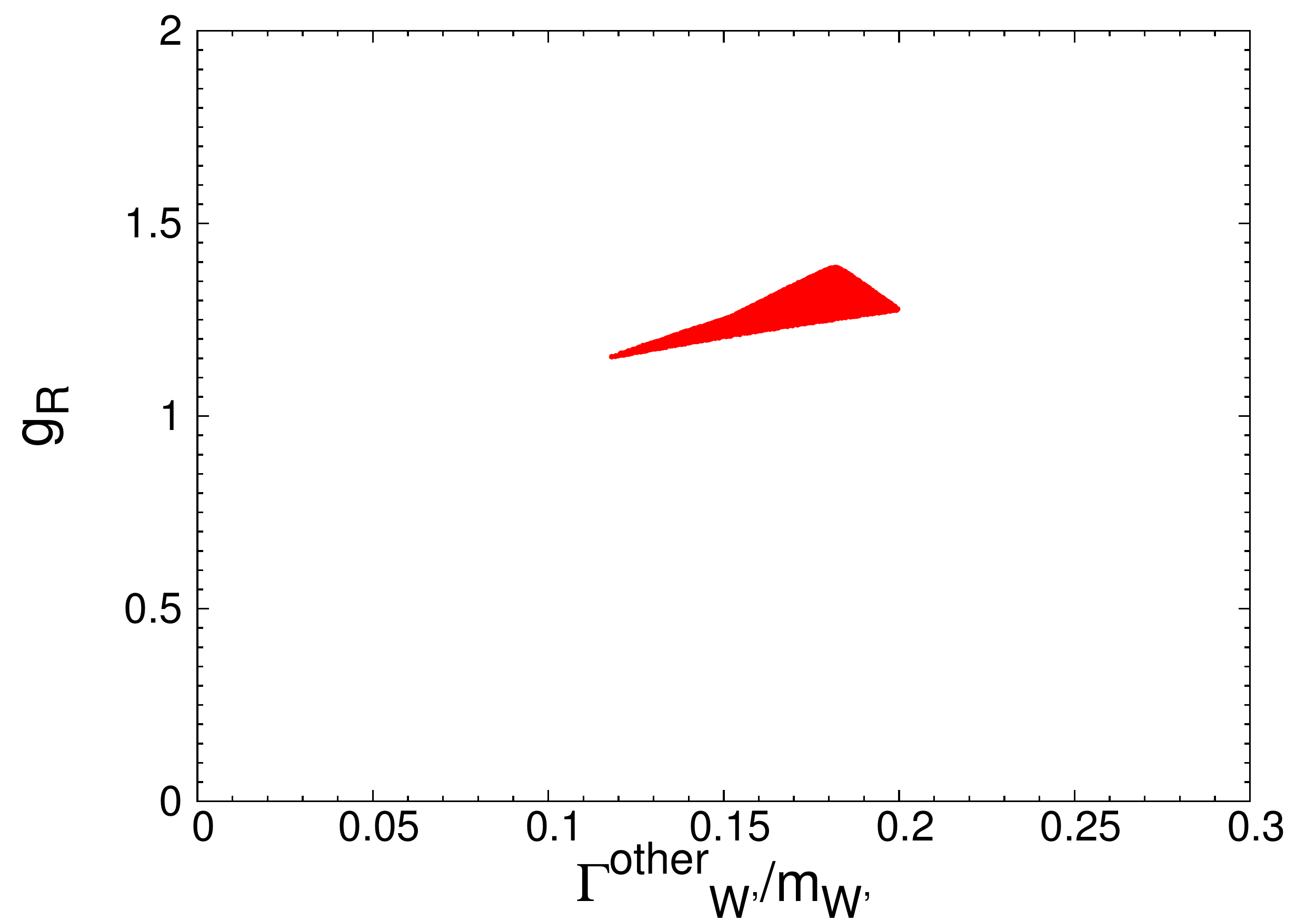}
\caption{\small \label{f1} 
The red-dotted points for $m_{W'}=3$
TeV satisfy the signal production cross section 
$\sigma(pp \to W' \to WH)\geq 4.5$ fb,
 the upper limits listed in Table~\ref{tab:constraint}, 
and the dijet upper limit adapted for the broad-width resonance. 
The cyan (green) hatched region was excluded by the Higgs precision data 
of Type-I 2HDM~\cite{atlas-2hdm,atlas_cms_data}.
}
\end{figure}

\begin{figure}[t!]
\centering
\includegraphics[height=4.2in,angle=-90]{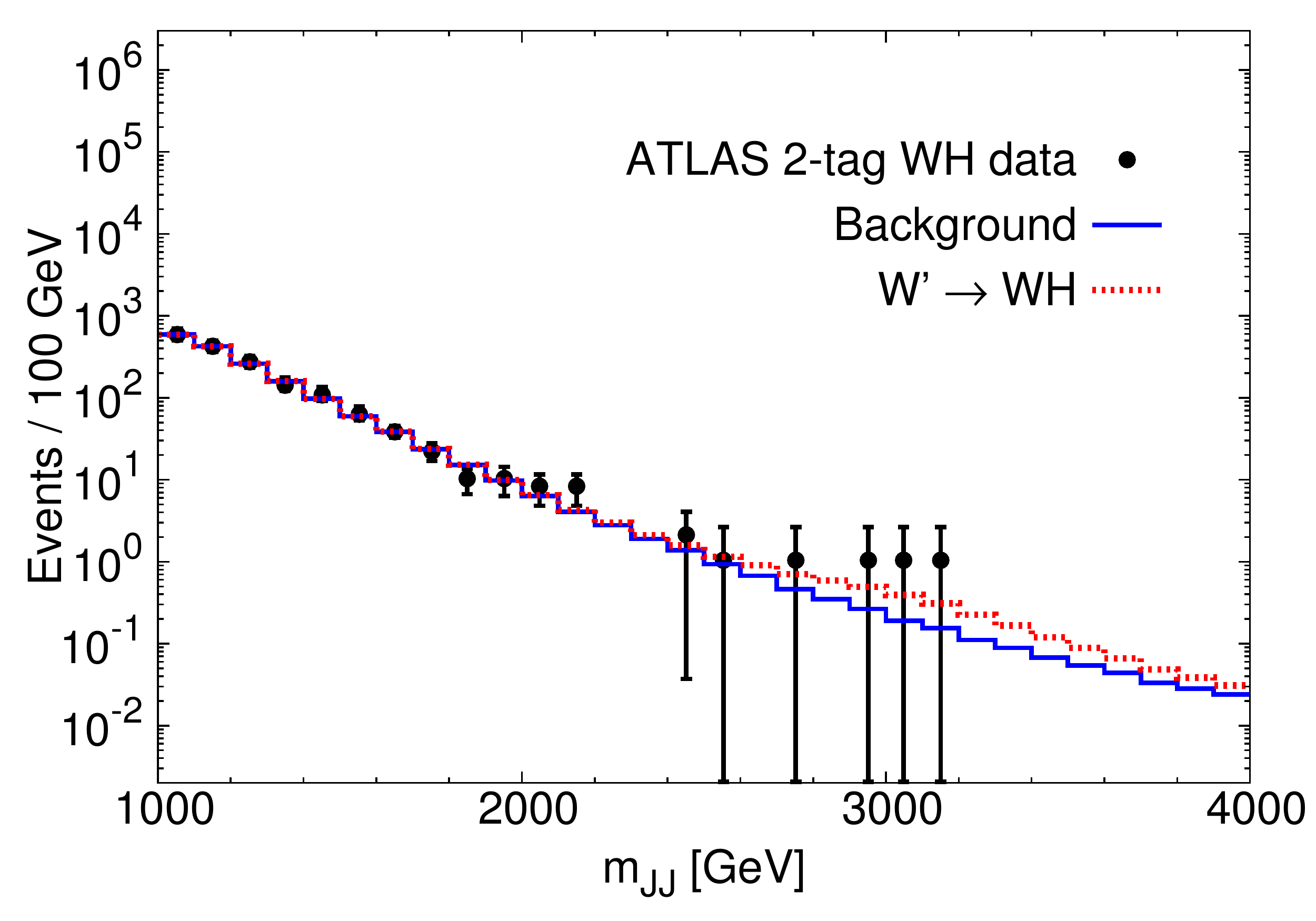}
\caption{\small \label{f2} 
The $m_{JJ}$ distribution of ATLAS 2-tag $WH$ data points 
and the SM background (blue-solid histogram) are from Ref.~\cite{atlas}.
The $W' \to WH$ contribution added to the background is indicated with 
the red-dashed histogram.
The parameters are $\cos(\beta-\alpha)=-0.3$, $\tan\beta=2.41$, $g_R=1.358$, 
$\Gamma^{\rm other}_{W'}=0.185\times m_{W'}$, and $m_{W'}=3$ TeV, which gives
$\sigma(pp \to W' \to WH)=9.7$ fb, $\Gamma_{W'}=0.3\times m_{W'}$.
}
\end{figure}

In Fig.~\ref{f1}, we show the aforementioned experimental constraints 
on the parameter space 
of $SU(2)_L\times SU(2)_R\times U(1)_{B-L}$ model, 
and include the non-standard $W'$ decay width $\Gamma^{\rm other}_{W'}$.
The red-dotted points satisfy the requirement on the signal cross section 
\begin{equation}
\label{x-sec45}
\sigma(pp \to W') \times B(W' \to WH) \geq 4.5 \; {\rm fb},
\end{equation}
evaluated in the narrow-width approximation,
and the upper limits listed in Table~\ref{tab:constraint}, 
except for the dijet upper limit. 
The dijet limits are adapted to the broad-width-resonance case, 
following the instructions in Ref.~\cite{ATLAS:dijet,Khachatryan:2015sja}. 
The excess bump in the $m_{JJ}$ distribution of the 2-tag $WH$ 
channel from ATLAS~\cite{atlas}
is not necessarily a narrow resonance, 
likewise, we do not restrict the width of $W'$ to be narrow.
%
%
%
The cyan (green)  hatched region was excluded by the combined 7 and 8 TeV
ATLAS and CMS signal strength data (the ATLAS data only)
~\cite{atlas-2hdm,atlas_cms_data}. 
The shifting of the hatched region is mainly due to the change in the
diphoton signal strength from $\mu_{\gamma\gamma}(ggF) = 1.32 \pm 0.38$
to $1.10 ^{+0.23}_{-0.22}$.  Most of the red-dotted points are ruled 
out by this constraint, yet there exists a small region that satisfies 
all the existing constraints and Higgs precision data.

Nevertheless, as shown in Fig~\ref{f1} 
there exists a small region of parameter space 
that is not excluded by the aforementioned constraints, including
all those listed in Table~\ref{tab:constraint} 
(with modified dijet constraints) and the Higgs precision data, as
well as satisfying the cross section requirement in Eq.~(\ref{x-sec45}).
This small region corresponds to parameters
$\cos(\beta-\alpha) \simeq  -0.3$, $\tan\beta \simeq 2.41$, $g_R \simeq 1.358$, 
and $\Gamma^{\rm other}_{W'} \simeq 0.185\times m_{W'}$.
It will give a cross section of 
$\sigma(pp \to W') \times B(W' \to WH ) \approx 4.6$ fb
in the narrow-width approximation.
However, if we abandon the narrow-width approximation and adopt the full
calculation, it gives a cross section of 
$\sigma(pp \to W' \to WH)=9.7$ fb and $\Gamma_{W'}=0.3\times m_{W'}$.
Thus, $\sigma(pp \to W' \to WH) \times B(H\to b \bar b) \approx 5.2 $ fb,
\footnote{
We employed the branching ratio $B(H \to b\bar b) =0.54$ in 2HDM type-I
for $\cos(\beta-\alpha)=-0.3$ and $\tan\beta = 2.41$.}
which is within the range shown by the ATLAS data in Eq.~(\ref{first}).
Note that the $K$ factor for the process is roughly $1.3$ at the LHC
energies, but for the purpose of consistency with backgrounds we do not
multiply this $K$ factor.
Using this point, the $W'$ contribution to the $m_{JJ}$ distribution is shown 
in Fig.~\ref{f2} with red-dashed histograms,  
where $m_{JJ}$ is the invariant mass of the $W$ and $H$ hadronic jets.
We can see that this broad-width $W'$ provides an interpretation for 
the three observed events around $m_{JJ}=3$ TeV of ATLAS.
Therefore, the allowed region, though small, can explain
the excess bump observed at the $WH$ channel.

Additional comments are in order here. From the upper-left panel 
in Fig.~\ref{f1}, the distribution of the red-dotted points is symmetric
under the exchange $\tan\beta \leftrightarrow 1/\tan\beta$ 
and $\cos(\beta-\alpha)\leftrightarrow -\cos(\beta-\alpha)$.
It is because  the ratio between $\Gamma(W' \to WH)$ and $\Gamma(W' \to WZ)$
can be rewritten as
\begin{equation}
\frac{\Gamma(W' \to WH)}{\Gamma(W' \to WZ)}=\frac{\cos(\beta+\alpha)}{\sin(2\beta)}
=\frac{1}{2}\left[ \frac{1}{\tan\beta}-\tan\beta \right]\times \cos(\beta-\alpha)+\sin(\beta-\alpha).
\end{equation}
Also, from the lower-right panel we can see that without the 
non-standard decay of $W'$, 
the $W'$ of $SU(2)_L\times SU(2)_R\times U(1)_{B-L}$ model 
does not have any more viable parameter space to 
explain the $WH$ excess observed at ATLAS,
mainly due to the dijet constraint.

\section{Conclusions}

We have studied a unified model based on $SU(2)_L \times SU(2)_R \times 
U(1)_{B-L}$,
which was broken at multi-TeV scale to the SM symmetry. We have 
attempted to use the $W'$ gauge boson of mass 3 TeV to interpret
the excess bump seen at the ATLAS $W H \to (q \bar q') (b\bar b)$ data. 
We have shown that such an interpretation faces very strong constraints
from dijet data and $WZ$ data, as well as the precision Higgs data.
Yet, we are able to find a viable parameter
space region, though small, that can accommodate all the existing data
and provide an explanation for the excess bump at 3 TeV.  The largest 
cross section that we obtain 
is $\sigma(pp \to W' \to WH) \times B(H \to b \bar b) \simeq 5.2 $ fb,
which is roughly equal to the experimental result shown in Eq.~(\ref{first}).

A few comments are offered as follows.
\begin{enumerate}

\item Below the symmetry breaking scale of 
$SU(2)_L \times SU(2)_R \times U(1)_{B-L} \to SU(2)_L \times U(1)_Y$, the Higgs 
field can be recast into two doublet Higgs fields, in a manner similar 
to the conventional 2HDM. Therefore, the model is also subject to the
constraints from the precision Higgs data.  The ATLAS publication
\cite{atlas-2hdm} has presented the excluded region in various 2HDM's.
We adopted the least restricted one -- Type I -- in this work, and showed
the excluded region in the upper-left panel Fig.~\ref{f1}.  All the
other types of 2HDM's are more severely constrained, and have no allowed
region when superimposed on our model.

\item 
The mass spectrum of $A$, $H^+$, and $H'$ will have interesting effects on 
flavor physics and low energy constraints.  First of all, $B$ physics
is sensitive to the charged Higgs mass, e.g., $b\to s \gamma$, 
$B$-$\overline{B}$ mixing, $B \to \mu^+ \mu^-$. 
However, in Type I 2HDM all Yukawa couplings are
proportional to $\cos\alpha/\sin\beta$.  Therefore, based on the constraint
from Higgs precision data, $\sin(\beta-\alpha) \ \approx \  1$, and so that
$\alpha \simeq \beta - \pi/2$. It implies that $\cos\alpha/\sin\beta
\simeq 1$.  Hence, there is no $\tan\beta$ enhancement in contrast to the 
Type II model. Therefore, as long as $\tan\beta \ ^>_\sim \  1$, 
the constraint on the charged Higgs mass is rather weak.
Another important constraint is the $\rho$ parameter (or $\Delta T$) being
very close to 1 -- the custodial limit.  It can be fulfilled by
taking the mass splitting among $A, H', H^+$ to be small.  We therefore set 
$m_A \approx m_{H'} \approx m_{H^+}$.

\item We have adopted the leptophobic condition for the $W'$ boson, or by
assuming the right-handed neutrino is heavier than the mass of $W'$.  

\item 
Note that the boson jets for $W$ and $Z$ bosons are overlapping at 60\%.
We do not work out for the $Z' \to ZH$ boson in this work, but it
can be done similarly.  However, leptophobic version is a must for 
the $Z'$ to avoid the very strong leptonic limit.  

\item 
The dijet limit of $pp \to W' \to jj$ presented the most stringent constraint
to the model.  We have to adopt other decay modes in order to dilute 
the branching ratio into dijets. Possible decay modes are
$  W'^+ \to H^+A,\;Z H^+,\; W^+ H',\; W^+ A,\; H^+ H,\; H^+ H'$. 
Searches for these modes serve as further checks on the model.

\item 
The ATLAS data (and also the CMS data) did not indicate a narrow resonance 
at 3 TeV. Therefore, we assume one more parameter (somewhat restricted)
$\Gamma^{\rm other}_{W'}$ to alleviate the constraint from dijet. As shown 
in Fig.~\ref{f2}, the resonance width is rather wide. Currently, we obtained
the total width $\Gamma_{W'} = 0.3 \times m_{W'}$.

\item
Although there are some direct searches on $A$ and $H'$ from the
LHC~\cite{2HDM:Collider}, the constraints for Type-I 2HDM are 
not strong enough. Conservatively, we can focus on heavy
Higgs bosons around $500 - 1000$ GeV, and the interesting signatures
for this mass range can be categorized according to their final
states:
\begin{eqnarray}
\hbox{I.} & \quad& 
p p\rightarrow W^{\prime +} \rightarrow H^+Z/H \rightarrow 
t\overline{b} b\overline{b}\, (\tau^+\tau^-) \rightarrow W^+ +4b\quad 
{\rm or} \quad W^+ +2b2\tau.  \ .\nonumber\\
\hbox{II.} &\quad&
p p\rightarrow W^{\prime +} \rightarrow H^+A/H' \rightarrow 
t\overline{b}t\overline{t} \rightarrow W^+W^+W^- +4b \;. \nonumber\\
\hbox{III.} &\quad&
p p\rightarrow W^{\prime +} \rightarrow W^+A/H' 
\rightarrow W^+t\overline{t} \rightarrow W^+W^+W^- +2b \;. \nonumber
\end{eqnarray}
In the second one, the $W^+ W^+ W^-$ can decay into a pair of 
same-sign dilepton and a pair of jets plus missing energy. Indeed, 
it has been searched for at the LHC~\cite{ss_dilepton_8tev}.
Many other possibilities of final states consisting of multi-leptons 
and jets can also be searched for.
All these channels are to be explored if the excess of the 
3 TeV $WH$ resonance is going to be established in the future data.

\end{enumerate}

\section*{\bf Acknowledgments}
This research was supported in parts 
by the MoST of Taiwan
under Grant No. MOST-105-2112-M-007-028-MY3,
by the U.S. DOE under Grant No. DE-FG-02-12ER41811 at UIC,
and by the World Premier International Research Center Initiative (WPI), 
MEXT, Japan.


\end{document}